\documentclass[prl,reprint,preprintnumbers,amsmath,amssymb,floatfix,aps,longbibliography]{revtex4-1}

\usepackage[utf8]{inputenc}
\usepackage[T1]{fontenc}
\usepackage{txfonts}
\usepackage{microtype}
\usepackage{eucal}
\usepackage{bm}
\usepackage{physics}
\usepackage{xr}
\usepackage{empheq}
\usepackage{lipsum}
\usepackage{subcaption}
\usepackage{color,verbatim}
\usepackage{soul}

\usepackage{graphicx}
\usepackage{float}
\usepackage[draft]{pgf}

\usepackage{booktabs}

\usepackage[colorlinks,allcolors=blue]{hyperref}
\usepackage[capitalize]{cleveref}

\newcommand{\up}{\uparrow}                                      
\newcommand{\dn}{\downarrow}                                    

\definecolor{Haakon}{rgb}{0,0,1}

\newcommand{\ve}[1]{\boldsymbol{#1}}

\newcommand{\nodag}{{\mathstrut}}

\begin{document}
\title{Magnon decay theory of Gilbert damping in metallic antiferromagnets}
\author{Haakon T. Simensen}
\author{Akashdeep Kamra}
\author{Roberto E. Troncoso}
\author{Arne Brataas}
\affiliation{Center for Quantum Spintronics, Department of Physics, Norwegian University of Science and Technology, NO-7491 Trondheim, Norway}

\date{\today}
 
\begin{abstract}
Gilbert damping is a key property governing magnetization dynamics in ordered magnets. We present a theoretical study of intrinsic Gilbert damping induced by magnon decay in antiferromagnetic metals through $s$-$d$ exchange interaction. Our theory delineates the qualitative features of damping in metallic antiferromagnets owing to their bipartite nature, in addition to providing analytic expressions for the damping parameters. Magnon-induced intraband electron scattering is found to predominantly cause magnetization damping, whereas the Néel field is found to be damped via disorder. Depending on the conduction electron band structure, we predict that magnon-induced interband electron scattering around band crossings may be exploited to engineer a strong Néel field damping.
\end{abstract}

\maketitle


\textit{Introduction.---}The dynamical properties of a harmonic mode are captured by its frequency and lifetime~\cite{Feynman1963,Mahan2000}. While the eigenfrequency is typically determined by the linearized equations of motion, or equivalently by a non-interacting description of the corresponding quantum excitation, the lifetime embodies rich physics stemming from its interaction with one or more dissipative baths~\cite{Feynman1963,Caldeira1981}. Dissipation plays a central role in the system response time. In the context of magnetic systems employed as memories, the switching times decrease with increasing damping thereby requiring a stronger dissipation for fast operation~\cite{Kimel2004,KirilyukRMP2010,Brataas2012}. The dissipative properties of the system also result in rich phenomena such as quantum phase transitions~\cite{Fazio2001,Biella2015,Maile2018,Rastelli2018}. Furthermore, the formation of hybrid excitations, such as magnon-polarons~\cite{Kittel1958,Weiler2012,Ruckriegel2014,Kamra2015,Flebus2017,Kikkawa2016,Dreher2012,Simensen2019} and magnon-polaritons~\cite{Huebl2013,Tabuchi2015,Hu2015,Johansen2018,Harder2018A,Cao2015}, requires the dissipation to be weak with respect to the coupling strengths between the two participating excitations~\cite{FriskKockum2019}. Therefore, in several physical phenomena that have emerged into focus in the recent years~\cite{Holanda2018,Weiler2012,Kikkawa2016,ViolaKusminskiy2016,Samkharadze2018,Harder2018,Herskind2009}, damping not only determines the system response but also the very nature of the eigenmodes themselves. Understanding, exploiting and controlling the damping in magnets is thus a foundational pillar of the field. 

The success of Landau-Lifshitz-Gilbert (LLG) phenomenology \cite{Landau1935, Gilbert2004} in describing ferromagnetic dynamics has inspired vigorous efforts towards obtaining the Gilbert damping parameter using a wide range of microscopic theories. The quantum particles corresponding to magnetization dynamics - magnons - provide one such avenue for microscopic theories and form the central theme in the field of magnonics~\cite{Kruglyak2010,Chumak2015}. While a vast amount of fruitful research has provided a good understanding of ferromagnets (FMs) \cite{Mitchell1957, Heinrich1967, Kambersky1970, Kambersky1976, Kunes2002, Dobin2003, Zhang2004, Tserkovnyak2004, Kambersky2007, Skadsem2007, Gilmore2007, Brataas2008, Hickey2009, Starikov2010, Mankovsky2013, Yuan2014, Barati2014, Mahfouzi2017, Zhao2018, Li2019}, analogous studies on antiferromagnets (AFMs) are relatively scarce and have just started appearing~\cite{Moriyama2019,Liu2017} due to the recently invigorated field of antiferromagnetic spintronics~\cite{Jungwirth2016,Baltz2018,MacDonald2011,Smejkal2018,Gomonay2014,Gomonay2018}. Among the ongoing discoveries of niches borne by AFMs, from electrically and rapidly switchable memories \cite{Wadley2016}, topological spintronics \cite{Smejkal2018}, long range magnonic transport \cite{LebrunNature2018} to quantum fluctuations \cite{Kamra2019}, an unexpected surprise has been encountered in the first principles evaluation of damping in metallic AFMs. Liu and coworkers \cite{Liu2017} and another more recent first-principles study \cite{Mahfouzi2018} both found the magnetization dissipation parameter to be much larger than the corresponding Néel damping constant, in stark contrast with previous assumptions, exhibiting richer features than in FMs. An understanding of this qualitative difference as well as the general AFM dissipation is crucial for the rapidly growing applications and fundamental novel phenomena based on AFMs.

Here, we accomplish an intuitive and general understanding of the Gilbert damping in metallic AFMs based on the magnon picture of AFM dynamics. Employing the $s$-$d$, two-sublattice model for a metallic AFM, in which the $d$ and $s$ electrons constitute the magnetic and conduction subsystems, we derive analytic expressions for the Gilbert damping parameters as a function of the conduction electron density of states at the Fermi energy and $s$-$d$ exchange strength. The presence of spin-degenerate conduction bands in AFMs is found to be the key in their qualitatively different damping properties as compared to FMs. This allows for absorption of AFM magnons via $s$-$d$ exchange-mediated intraband conduction electron spin-flip processes leading to strong damping of the magnetization as compared to the Néel field~\footnote{Such processes are forbidden in ferromagnets due to spin splitting between the opposite spin conduction electron bands.}. We also show that interband spin-flip processes, which are forbidden in our simple AFM model but possible in AFMs with band crossings in the conduction electron dispersion, result in a strong Néel field damping. Thus, the general qualitative features of damping in metallic AFMs demonstrated herein allow us to understand the Gilbert damping given the conduction electron band structure. These insights provide guidance for engineering AFMs with desired damping properties, which depend on the exact role of the AFM in a device.

\begin{figure}[tb]
\centering
\includegraphics[width=1.0\columnwidth]{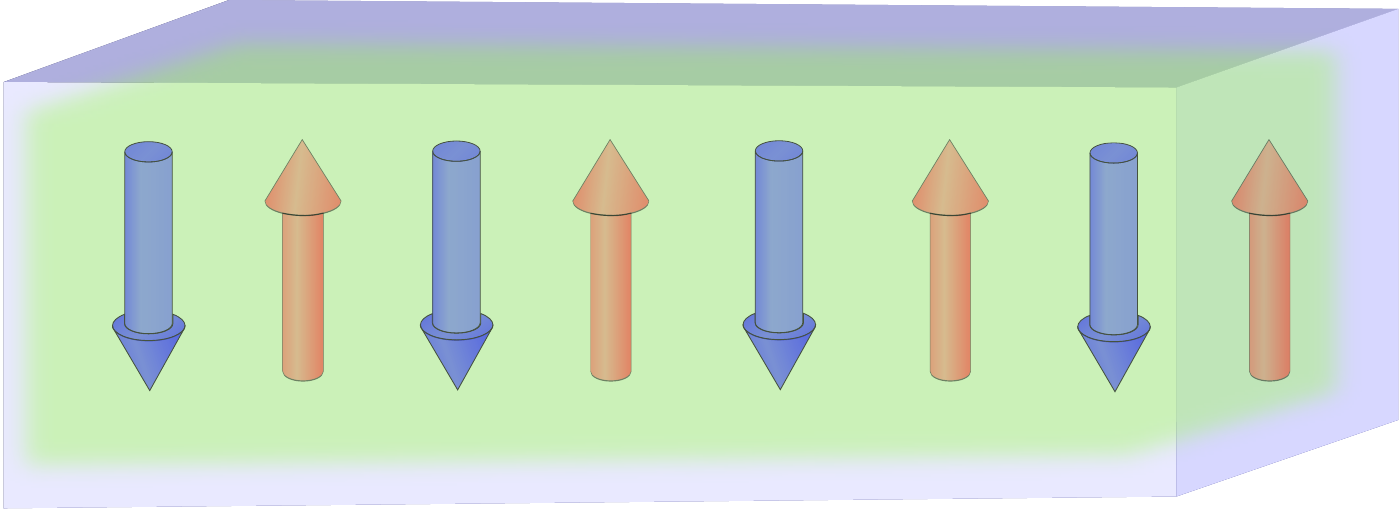}
\caption{Schematic depiction of our model for a metallic AFM. The red and blue arrows represent the localized $d$ electrons with spin up and down, respectively, thereby constituting the Néel ordered magnetic subsystem. The green cloud illustrates the delocalized, itinerant $s$ electrons that forms the conduction subsystem.}
\label{fig:model}
\end{figure}

\textit{Model.---}We consider two-sublattice metallic AFMs within the $s$-$d$ model \cite{Mitchell1957,Heinrich1967,Skadsem2007}. The $d$ electrons localized at lattices sites constitute the magnetic subsystem responsible for antiferromagnetism, while the itinerant $s$ electrons form the conduction subsystem that accounts for the metallic traits. The two subsystems interact via $s$-$d$ exchange [Eq. (\ref{eq:sdex})]. For ease of depiction and enabling an understanding of qualitative trends, we here consider a one-dimensional AFM (Fig. \ref{fig:model}). The results within this simple model are generalized to AFMs with any dimensionality in a straightforward manner. Furthermore, we primarily focus on the uniform magnetization dynamics modes.

At each lattice site $i$, there is a localized $d$ electron with spin $\ve{S}_i$. The ensuing magnetic subsystem is antiferromagnetically ordered (Fig. \ref{fig:model}), and the quantized excitations are magnons \cite{Anderson1952,Kubo1952}. Disregarding applied fields for simplicity and assuming an easy-axis anisotropy along the $z$-axis, the magnetic Hamiltonian, $\mathcal{H}_{\rm m} = \tilde{J} \sum_{\langle i, j \rangle} \ve{S}_i \cdot \ve{S}_j - K \sum_i (S_i^z)^2$, where $\langle i,j \rangle$ denotes summation over nearest neighbor lattice sites, is quantized and mapped to the sublattice-magnon basis \cite{Kubo1952}
\begin{equation}\label{eq:ham1}
    \mathcal{H}_{\rm m} = \sum_{\ve{q}} \left[  A_{\ve{q}} \left( a_{\ve{q}}^\dagger a_{\ve{q}}^\nodag + 
    b_{\ve{q}}^\dagger b_{\ve{q}}^\nodag  \right)  + 
      B_{\ve{q}}^\dagger a_{\ve{q}}^\dagger b_{\ve{q}}^\dagger + B_{\ve{q}} a_{\ve{q}}^\nodag b_{\ve{q}}^\nodag   \right],
\end{equation}

\noindent where we substitute $\hbar = 1$, $A_{\ve{q}} = (2\tilde{J} + 2K) S$ and $B_{\ve{q}} = \tilde{J} S e^{-i \ve{q} \cdot \ve{a}} \sum_{\langle \ve{\delta} \rangle} e^{i \ve{q} \cdot \ve{\delta}}$, where $S = |\ve{S}_i|$, $\ve{a}$ is the displacement between the two atoms in the basis, and $\langle \ve{\delta} \rangle$ denotes summing over nearest neighbor displacement vectors. $a_{\ve{q}}^{\nodag}$ and $b_{\ve{q}}^{\nodag}$ are bosonic annihilation operators for plane wave magnons on the A and B sublattices, respectively.  We diagonalize the Hamiltonian [Eq. \ref{eq:ham1}] through a Bogoliubov transformation \cite{Kubo1952} to $\mathcal{H}_{\rm m} = \sum_{\ve{q}} \omega_{\ve{q}} \left( \alpha_{\ve{q}}^\dagger \alpha_{\ve{q}}^\nodag + 
\beta_{\ve{q}}^\dagger \beta_{\ve{q}}^\nodag  \right),$ with eigenenergies  $\omega_{\ve{q}} = \sqrt{A_{\ve{q}}^2 - |B_{\ve{q}}|^2}$. In the absence of an applied field, the magnon modes are degenerate.

\begin{figure}[tb]
\centering
\includegraphics[width=1\linewidth]{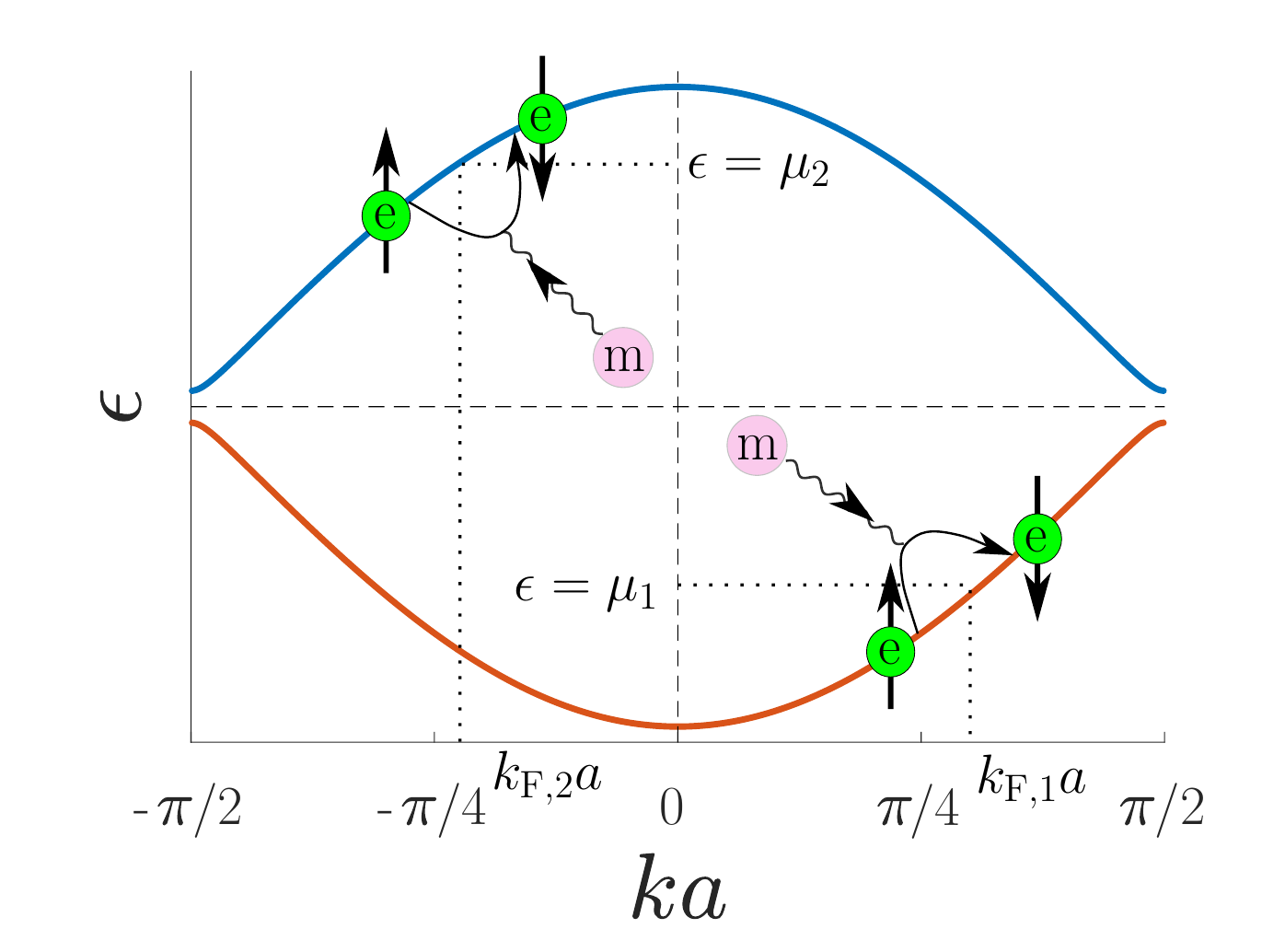}
\caption{The $s$ electron dispersion in metallic AFM model, where the red and blue dispersions depict electron bands 1 and 2, respectively. Illustrations of intraband electron-magnon scattering at two different Fermi levels, $\mu_1$ and $\mu_2$, are added. The depicted momentum transfer is exaggerated for clarity.}
\label{fig:disp}
\end{figure}

The $s$ electron conduction subsystem is described by a tight-binding Hamiltonian that includes the ``static'' contribution from the $s$-$d$ exchange interaction [Eq. (\ref{eq:sdex})] discussed below:
\begin{equation}
    \mathcal{H}_{\rm e}  = -t \sum_{\langle i,j \rangle} \sum_{\sigma}  c_{i \sigma}^\dagger c_{j \sigma}^\nodag - J \sum_{i} (-1)^i \left( c_{i \up}^\dagger c_{i \up}^\nodag - c_{i \dn}^\dagger c_{i \dn}^\nodag \right).
    \label{eq:it_electron}
\end{equation}
Here $c_{i \sigma}^{\nodag}$ is the annihilation operator for an $s$ electron at site $i$ with spin $\sigma$. $t~(> 0)$ is the hopping parameter, and $J~(> 0)$ accounts for $s$-$d$ exchange interaction [Eq. (\ref{eq:sdex})]. The $(-1)^i$ factor in the exchange term reflects the two-sublattice nature of the AFM. The conduction subsystem unit cell consists of two basis atoms, similar to the magnetic subsystem. As a result, there are four distinct electron bands: two due to there being two basis atoms per unit cell, and twice this due to the two possible spin polarizations per electron. Disregarding applied fields, these bands constitute two spin-degenerate bands. We label these bands 1 and 2, where the latter is higher in energy. The itinerant electron Hamiltonian [Eq. \eqref{eq:it_electron}] is diagonalized into an eigenbasis $(c_{1 \ve{k} \sigma}, c_{2 \ve{k} \sigma})$ with eigenenergies $\epsilon_{1 \ve{k}} = -\epsilon_{\ve{k}}$ and $\epsilon_{2 \ve{k}} = +\epsilon_{\ve{k}}$, where $\epsilon_{\ve{k}} =  \sqrt{ J^2 S^2 + t^2 |\gamma_{\ve{k}}|^2 }$, where $\gamma_{\ve{k}} = \sum_{\langle \ve{\delta} \rangle} e^{-i \ve{k} \cdot \ve{\delta}}$. The itinerant electron dispersion is depicted in Fig. \ref{fig:disp}.

The magnetic and conduction subsystems interact through $s$-$d$ exchange interaction, parametrized by $J$:
\begin{equation}\label{eq:sdex}
    \mathcal{H}_{\rm I} = - J \sum_{i} \ve{S}_i \cdot \ve{s}_i,
\end{equation}
where $\ve{s}_i = \sum_{\sigma \sigma'} c_{i \sigma}^\dagger \boldsymbol{\sigma}_{\sigma \sigma'} c_{i \sigma'}$ is the spin of the itinerant electrons at site $i$, where $\boldsymbol{\sigma}$ is the vector of Pauli matrices. The term which is zeroth order in the magnon operators, and thus accounts for the static magnetic texture, is already included in $\mathcal{H}_{\rm e}$ [Eq. \eqref{eq:it_electron}]. To first order in magnon operators, the interaction Hamiltonian can be compactly written as
\begin{align}
    \begin{split}
    \mathcal{H}_{\rm e-m} &=
         \sum_{\substack{\lambda \rho }} \sum_{ \ve{k} \ve{k}' \ve{q}} 
         c_{\lambda \ve{k} \up}^\dagger c_{\rho \ve{k}' \dn}^\nodag \left( W^{\mathrm{A}, \lambda \rho}_{\ve{k}\ve{k}'\ve{q}} a_{-\ve{q}}^\dagger +  W^{\mathrm{B}, \lambda \rho}_{\ve{k}\ve{k}'\ve{q}} b_{\ve{q}}^\nodag \right)
         +\,  \textrm{h.c.},
    \end{split}
    \label{eq:HI}
\end{align}
\noindent where $\lambda$ and $\rho$ are summed over the electron band indices. As detailed in the Supplemental material, $W^{\mathrm{A}, \lambda \rho}_{\ve{k}\ve{k}'\ve{q}}$ and $W^{\mathrm{B}, \lambda \rho}_{\ve{k}\ve{k}'\ve{q}}$, both linear in $J$, are coefficients determining the amplitudes for scattering between the itinerant electrons and the $a_{\ve{q}}^\nodag$ and $b_{\ve{q}}^\nodag$ magnons, respectively. Specifically, when considering plane wave states, $W^{\mathrm{A}/\mathrm{B}, \lambda \rho}_{\ve{k}\ve{k}'\ve{q}}$ becomes a delta function, thereby enforcing the conservation of crystal momentum in a translationally invariant lattice. Inclusion of disorder or other many-body effects results in deviation of the eigenstates from ideal plane waves causing a wave vector spread around its mean value \cite{Mahan2000}. The delta function, associated with an exact crystal momentum conservation, is thus transformed to a peaked function with finite width $(\Delta k)$. The $\lambda\rho$ combinations $11$ and $22$ describe \textit{intraband} electron scattering, while $12$ and $21$ describe \textit{interband} scattering. Intraband scattering is illustrated in Fig. \ref{fig:disp}. Interband scattering is prohibited within our model due to energy conservation, since the uniform $\ve{q} = \ve{0}$ magnon energy is much smaller than the band gap.

The scattering described by $\mathcal{H}_{\rm e-m}$ [Eq. \eqref{eq:HI}] transfers spin angular momentum between the magnetic and conduction subsystems. The itinerant electrons are assumed to maintain a thermal distribution thereby acting as a perfect spin sink. This is consistent with a strong conduction electron spin relaxation observed in metallic AFMs \cite{Merodio2014, Frangou2016}. As a result, the magnetic subsystem spin is effectively damped through the $s$-$d$ exchange interaction.

\textit{Gilbert damping.---}In the Landau-Lifshitz-Gilbert (LLG) phenomenology for two-sublattice AFMs, dissipation is accounted via a 2$\times$2 Gilbert damping matrix~\cite{Kamra2018}. Our goal here is to determine the elements of this matrix in terms of the parameters and physical observables within our microscopic model. To this end, we evaluate the spin current ``pumped'' by the magnetic subsystem into the $s$ conduction electrons, which dissipate it immediately within our model. The angular momentum thus lost by the magnetic subsystem appears as Gilbert damping in its dynamical equations~\cite{Kamra2018,Kamra2016A}. The second essential ingredient in identifying the Gilbert damping matrix from our microscopic theory is the idea of coherent states~\cite{Glauber1963,Sudarshan1963}. The classical LLG description of the magnetization is necessarily equivalent to our quantum formalism, when the magnetic eigenmode is in a coherent state~\cite{Glauber1963,Sudarshan1963,Zagury1971}. Driving the magnetization dynamics via a microwave field, such as in the case of ferromagnetic resonance experiments, achieves such a coherent magnetization dynamics~\cite{Kamra2016A,Kamra2016}.

The spin current pumped by a two-sublattice magnetic system into an electronic bath may be expressed as~\cite{Kamra2017}
\begin{align}
\begin{split}
    I_{z}   &= G_{mm} \left( \ve{m} \times \dot{\ve{m}}\right)^z + G_{nn} \left( \ve{n} \times \dot{\ve{n}}\right)^z \\
    &\quad + G_{mn} \left[ \left( \ve{m} \times \dot{\ve{n}} \right)^z + \left( \ve{n} \times \dot{\ve{m}} \right)^z \right],
\end{split}
\label{eq:I2}
\end{align}
where $\ve{m}$ and $\ve{n}$ are the magnetization and Néel field normalized by the sublattice magnetization, respectively. Here, ${G_{ij}  = \alpha_{ij} \times \left(M/|\gamma| \right)}$, where $\alpha_{ij}$ are the Gilbert damping coefficients, $\gamma$ is the gyromagnetic ratio of the $d$ electrons and $M$ is the sublattice magnetization. Considering the uniform magnetization mode, $I_z$ is the spin current operator $I_z = i \comm{\mathcal{H}_{\rm e-m}}{S^z}$~\cite{Bender2015}, where $S^z = \sum_i S_i^z$. We get
\begin{equation}\label{eq:scop}
    I_z = i \sum_{\substack{\lambda \rho }} \sum_{ \ve{k} \ve{k}' \ve{q}} 
         c_{\lambda \ve{k} \up}^\dagger c_{\rho \ve{k}' \dn}^\nodag \left( W^{\mathrm{A}, \lambda \rho}_{\ve{k}\ve{k}'\ve{q}} a_{-\ve{q}}^\dagger +  W^{\mathrm{B}, \lambda \rho}_{\ve{k}\ve{k}'\ve{q}} b_{\ve{q}}^\nodag \right)
         -\,  \textrm{h.c.}.
\end{equation}
The expectation value of this operator assuming the uniform magnetization mode to be in a coherent state corresponds to the spin pumping current [Eq. (\ref{eq:I2})].

In order to evaluate the spin pumping current from Eq. (\ref{eq:scop}), we follow the method employed to calculate interfacial spin pumping current into normal metals in Refs. \cite{Kamra2016A,Kamra2016,Kamra2017}, and the procedure is described in detail therein. Briefly, this method entails assuming the magnetic and conduction subsystems to be independent and in equilibrium at $t = -\infty$, when the mutual interaction [Eq. \eqref{eq:HI}] is turned on. The subsequent time evolution of the coupled system allows evaluating its physical observables in steady state. The resulting coherent spin-current corresponds to the classical spin current $I_z$ that can be related to the motion of the magnetization and the Néel field [Eq. \eqref{eq:I2}]. As a last step, we identify expressions for $\left( \ve{m} \times \dot{\ve{m}}\right)^z$, $\left( \ve{m} \times \dot{\ve{n}}\right)^z$ and $\left( \ve{n} \times \dot{\ve{n}}\right)^z$ in terms of coherent magnon states, which enables us to identify the Gilbert damping coefficients $\alpha_{mm}$, $\alpha_{nn}$ and $\alpha_{mn}$.

\textit{Results.---}Relegating the detailed evaluation to Supplemental Material, we now present the analytic expression obtained for the various coefficients [Eq. (\ref{eq:I2})]. A key assumption that allows these simple expressions is that the electronic density of states in the conduction subsystem does not vary significantly over the magnon energy scale. Furthermore, we account for a weak disorder phenomenologically via a finite scattering length $l$ associated with the conduction electrons. This results in an effective broadening of the electron wavevectors determined by the inverse electron scattering length, $(\Delta k) = 2\pi/l$. As a result, the crystal momentum conservation in the system is enforced only within the wavevector broadening. By weak disorder we mean that the electron scattering length is much larger than the lattice parameter $a$. If $k$ and $k'$ are the wave vectors of the incoming and outgoing electrons, respectively, we then have $(k - k')a = (\Delta k) a \ll 1$. This justifies an expansion in the wave vector broadening $(\Delta k )a$. The Gilbert damping coefficients stemming from intraband  electron scattering are found to be 
\begin{align}
    \begin{split}
        \alpha_{mm} &= 
        \alpha_0(\xi_J) - \frac{\alpha_0(\xi_J)}{4}\left(1 + \frac{ \xi_J^2 \left( \xi_J^2 + 8 - 4 \cos^2(k_{\rm F} a) \right)}{\left(\xi_J^2 + 4 \cos^2(k_{\rm F} a)\right)^2} \right) \left[(\Delta k) a\right]^2,  \\
         \alpha_{nn} &= 
         \frac{\alpha_0(\xi_J)}{4} \left( 1 +
          \frac{\sin^2(k_{\rm F} a)}{\cos^2(k_{\rm F} a)} \frac{ \xi_J^2}{\left( \xi_J^2 + 4 \cos^2(k_{\rm F} a) \right)} \right) \left[(\Delta k) a\right]^2
        .
    \end{split}
    \label{eq:intra}
\end{align}
where $\xi_J = J S / t$, $k_{\rm F}$ is the Fermi momentum and $a$ is the lattice parameter, and where
\begin{equation}
    \alpha_0(\xi_J) = \frac{\pi v^2 J^2}{8} g^2(\mu) |\tilde{V}|^2 \frac{4 \cos^2(k_{\rm F} a)}{\xi_J^2 + 4 \cos^2(k_{\rm F} a)}.\end{equation}
Here, $v$ is the unit cell volume, $g(\epsilon)$ is the density of states per unit volume, $\mu$ is the Fermi level, and $\omega_0$ is the energy of the $\ve{q} = \ve{0}$ magnon mode. $\tilde{V}$ is a dimensionless and generally complex function introduced to account for the momentum broadening dependency of the scattering amplitudes. It satisfies $\tilde{V}(0) = 1$ and $0 \leq |\tilde{V}(\Delta k)| \leq 1$ within our model. These analytic expressions for the Gilbert damping parameters constitute one of the main results of this letter.

\textit{Discussion.--}We straightaway note that $\alpha_{nn}/\alpha_{mm} \sim \left[(\Delta k) a\right]^2 \ll 1$. $\alpha_{nn}$ is strictly dependent upon $(\Delta k) a$, and is non-zero only if there is disorder and a finite electron momentum broadening. $\alpha_{mm}$ is large even when considering a perfectly ordered crystal. This latter result is in good accordance with recent first-principles calculations in metallic AFMs \cite{Liu2017, Mahfouzi2018}. We moreover observe that both $\alpha_{mm}$ and $\alpha_{nn}$ are quadratic in $J$ and $g(\mu)$. This result is shared by Gilbert damping owing to spin-pumping in insulating ferrimagnet|normal metal (NM) and AFM|NM bilayers with interfacial exchange coupling~\cite{Kamra2017}. Metallic AFMs bear a close resemblance to these bilayer structures. There are however two main differences: The $s$-$d$ exchange coupling exists in the bulk of metallic AFMs, whereas it is localized at the interface in the bilayer structures. Additionally, the itinerant electron wave functions are qualitatively different in metallic AFMs and NMs, owing to the magnetic unit cell of the AFM. Indeed, these differences turn out to leave prominent signatures in the Gilbert damping in metallic AFMs.

\begin{figure}[tb]
\centering
\includegraphics[width=1\linewidth]{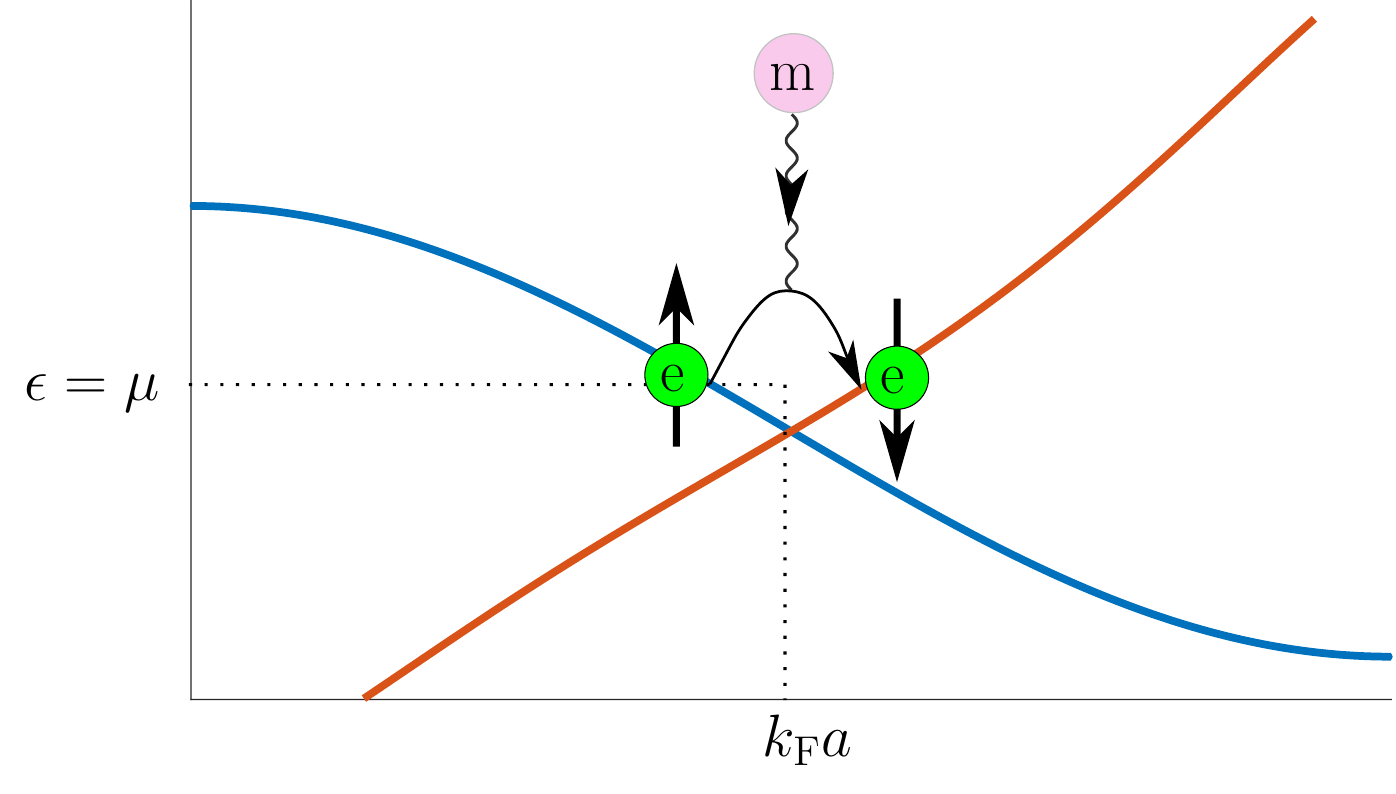}
\caption{A schematic depiction of magnon-induced interband scattering in a band crossing at the Fermi level.}
\label{fig:disp2}
\end{figure}

The uniform mode magnon energy is much smaller than the electron band gap within our simple model. Interband scattering is thus prohibited by energy conservation. However, in real AFM metals, the electron band structure is more complex. There may for instance exist band crossings \cite{Tang2016,Bodnar2018,Shao2019}. In such materials, magnon-induced interband electron scattering should also contribute to Gilbert damping, depending on the position of the Fermi surface. Motivated by this, we now consider Gilbert damping stemming from interband scattering, while disregarding the energy conservation for the moment, labeling the coefficients $\alpha_{mm}^I$ and $\alpha_{nn}^I$. We then find the same expressions as in Eq. (\ref{eq:intra}) with the roles of $\alpha_{mm,nn}^I$ interchanged with respect to $\alpha_{mm,nn}$. This implies that $\alpha_{nn}^I$ is large and independent of electron momentum broadening, whereas $\alpha_{mm}^I$ is proportional to the electron momentum broadening squared. Although arriving at this result required disregarding the energy conservation constraint, the qualitative effect in itself is not an artifact of this assumption. In materials with a band crossing, as depicted in Fig. \ref{fig:disp2},  $\alpha_{nn}^I/\alpha_{mm}^I > \alpha_{nn}/\alpha_{mm}$ is a general result. This generic principle derived within our simple model provides valuable guidance for designing materials with an engineered Gilbert damping matrix.

We now provide a rough intuitive picture for the damping dependencies obtained above followed by a more mathematical discussion. Consider a conventional diffraction experiment where an incident probing wave is able to resolve the two slits only when the wavelength is comparable to the physical separation between the two slits. In the case at hand, the wavefunctions of electrons and magnon participating in a scattering process combine in a way that the wavenumber by which the conservation of crystal momentum is violated becomes the probing wavenumber within a diffraction picture. Therefore, the processes conserving crystal momentum have vanishing probing wavenumber and are not able to resolve the opposite spins localized at adjacent lattice sites. Therefore, only the average magnetization is damped leaving the Néel field unaffected. With disorder, the probing wavenumber becomes non-zero and thus also couples to the Néel field. The interband scattering, on the other hand, is reminiscent of Umklapp scattering in a single-sublattice model and the probing wavenumber matches with the inverse lattice spacing. Therefore, the coupling with the Néel field is strong.  

The Gilbert damping in metallic AFMs here considered is caused by spin pumping from the magnetic subsystem into the $s$ band. This spin pumping induces electron transitions between spin $\up$/$\dn$ states among the $s$ electrons. The Gilbert damping coefficients depend thus on transition amplitudes proportional to products of itinerant electron wave functions such as $\psi_{\lambda k \up}^\dagger(x) \psi_{\rho k' \dn}^\nodag(x)$. The damping effect on sublattice A depends on this transition amplitude evaluated on the A sublattice, and equivalently for the B sublattice. Assuming without loss of generality that site $i = 0$ belongs to sublattice A, we find in the one-dimensional model that the damping on sublattice A is a function of $\sum_j \cos^2\left(\frac{\pi x_j}{2 a}\right) \psi_{\lambda k \up}^\dagger(x_j) \psi_{\rho k' \dn}^\nodag(x_j)$, whereas the damping on sublattice B is a function of $\sum_j \sin^2\left(\frac{\pi x_j}{2 a}\right) \psi_{\lambda k \up}^\dagger(x_j) \psi_{\rho k' \dn}^\nodag(x_j)$. Equivalently, by straightforwardly using that $m = (m_{\rm A} + m_{\rm B})/2$ and $n = (m_{\rm A} - m_{\rm B})/2$, this analysis predicts that $\alpha_{mm}$ is a function of $\sum_j \psi_{\lambda k \up}^\dagger(x_j) \psi_{\rho k' \dn}^\nodag(x)$, whereas $\alpha_{nn}$ is a function of $\sum_j \cos\left(\frac{\pi x_j}{a}\right) \psi_{\lambda k \up}^\dagger(x_j) \psi_{\rho k' \dn}^\nodag(x)$. Assuming plane wave solutions of the electron wave functions, and if we consider intraband scattering only, we more concretely find that $\alpha_{mm}$ is a function of $\left(1 - i (\Delta k) a \right)$, where $i$ is the imaginary unit, whereas $\alpha_{nn}$ is a function of $(\Delta k) a$. This coincides well with Eq. \eqref{eq:intra}.

Above, we presented a discussion of interband scattering in the minimal model where the band gap artificially was set to zero. In this limit, the upper electron band is a continuation of the lower band with a $\pm \pi/a$ momentum shift. We may then write $\psi_{2 k \sigma}^\nodag = \psi_{1, k + \pi/a, \sigma}^\nodag$. Under the assumption of a disappearing band gap, momentum-conserving interband scattering at momentum $k$ is therefore equivalent to intraband scattering between $k$ and $k \pm \pi/a$. This is the exact phase shift which results in a small $\alpha_{mm}$ and a large $\alpha_{nn}$ consistent with the discussion above. In real metallic AFMs with complex band structures, the exact wave function relations unveiled above do not apply. However, interband transition amplitudes will undoubtedly carry a position dependent phase. This position dependence results in a dephasing of transition amplitudes at neighboring lattice sites, which gives rise to a non-negligible $\alpha_{nn}$. The precise damping coefficients in real metallic AFMs depend on the detailed electron wave functions. We may however generally conclude that $\alpha_{nn}^I/\alpha_{mm}^I > \alpha_{nn}/\alpha_{mm}$.

\textit{Conclusion.---}We have provided a microscopic derivation of Gilbert damping resulting from magnon decay through $s$-$d$ exchange interaction in metallic antiferromagnets. Analytic expressions for Gilbert damping coefficients resulting from intraband electron scattering are presented, while Gilbert damping resulting from interband electron scattering is discussed on a conceptual level. We find that intraband electron scattering gives rise to a large magnetization damping and a negligible Néel field damping. The intraband Néel field damping is proportional to the inverse electron scattering length squared, and disappears exactly if there is no crystal disorder. By relating Gilbert damping to the degree to which transition amplitudes of the itinerant electron are in phase at neighboring lattice sites, we have argued for why interband electron scattering may generate a large Néel field damping.

\textit{Acknowledgments.---}This work was supported by the Research Council of
Norway through its Centres of Excellence funding scheme,
Project No. 262633 "QuSpin", the European Union’s Horizon 2020 Research and Innovation Programme under Grant DLV-737038 ”TRANSPIRE”, as well as by the European Research
Council via Advanced Grant No. 669442 "Insulatronics".
\renewcommand{\refname}{Bibliography}
%

\end{document}